\documentclass[reprint,
superscriptaddress,
bibnotes,
amsmath, amssymb,
aps,
prb,
floatfix]{revtex4-2}

\usepackage{tabularx}
\usepackage[table]{xcolor}
\usepackage{graphicx}%
\usepackage{dcolumn}%
\usepackage{bm}%
\usepackage{enumitem} %
\usepackage[colorlinks=true, allcolors=blue]{hyperref}%

\usepackage[T1]{fontenc}

\graphicspath{{figures/}}
\usepackage[separate-uncertainty=false]{siunitx}
\usepackage{braket}

\usepackage[capitalise]{cleveref}

\begin{document}

\title{Condensation dynamics in a two-dimensional photonic crystal waveguide}
\date{\today}

\author{Maria Efthymiou-Tsironi}
\thanks{These two authors contributed equally.}
\affiliation{CNR Nanotec, Institute of Nanotechnology, via Monteroni, 73100, Lecce, Italy}
\affiliation{Department of Physics, Universit\'a del Salento, via Monteroni, 73100, Lecce, Italy}

\author{Antonio Gianfrate}
\thanks{These two authors contributed equally.}
\author{Dimitrios Trypogeorgos}
\email[]{dimitrios.trypogeorgos@nanotec.cnr.it}
\author{Charly Leblanc}
\affiliation{CNR Nanotec, Institute of Nanotechnology, via Monteroni, 73100, Lecce, Italy}

\author{Fabrizio Riminucci}
\affiliation{Molecular Foundry, Lawrence Berkeley National Laboratory, One Cyclotron Road, Berkeley, California, 94720, USA}

\author{Grazia Salerno}
\affiliation{Department of Applied Physics, Aalto University School of Science, FI-00076 Aalto, Finland}

\author{Milena De Giorgi}
\author{Dario Ballarini}
\author{Daniele Sanvitto}
\affiliation{CNR Nanotec, Institute of Nanotechnology, via Monteroni, 73100, Lecce, Italy}

\begin{abstract}
Exciton-polariton condensation occurs at the extrema of the underlying dispersion where the density of states diverges and carriers can naturally accumulate.
The existence of multiple such points leads to coupling and competition between the associated modes and dynamical redistribution of the carriers in the dispersion.
Here, we directly engineer the above situation via subwavelength periodic patterning of a two-dimensional nanostructure.
This leads to multimode condensation into a pair of symmetric condensates that form at high-momenta, accidental-coupling points, and a high-symmetry $\Gamma$-point with a bound-in-the-continuum (BiC) state.
The dynamical behaviour of the system reveals the non-simultaneous appearance of these condensates and the interplay of non-trivial gain and relaxation mechanisms.
We fully characterise the quasi-static and dynamical regime of this artificial crystal and the properties of the different condensates.
This understanding is necessary when band-structure engineering techniques are used to achieve precise control of condensate formation with given energy and momentum.
\end{abstract}

\keywords{off-$\Gamma$ condensation, two-dimensional lattice, exciton-polaritons, BiC}

\maketitle

\section{Introduction}

Exciton-polaritons, arising from the strong coupling between light and matter in semiconductor quantum wells embedded in planar microcavity structures, have attracted significant interest due to their unique properties as degenerate Bose gases. 
These hybrid light-matter quasiparticles resulting from the strong coupling of excitons and photons, exhibit unique properties such as macroscopic quantum coherence and strong nonlinearities. 
These properties make exciton-polariton condensates an attractive platform for investigating fundamental aspects of Bose-Einstein condensation (BEC) and exploring innovative applications in optoelectronics and quantum information processing~\cite{carusotto2013}, even at room temperature~\cite{dusel2020,lerario2017,ghosh2022}.
Under certain conditions, these hybrid quasi-particles exhibit a dynamic phase transition characterised by the accumulation of a macroscopic number of particles in the lowest-energy, single-particle state with long-range order~\cite{kasprzak2006,amo2009a,amo2009b}. 
The light effective mass of exciton-polaritons enables them to achieve phase transition temperatures that are several orders of magnitude higher than those observed in atomic gases~\cite{anderson1995}.

Significant strides in our understanding of the condensation dynamics of exciton-polariton condensates have been made in the last few years, aided by the direct optical access of their coherence properties in both coordinate and momentum spaces.
It has been demonstrated that exciton-polariton condensation is a convenient way to obtain a laser, without the need for population inversion~\cite{imamog-lu1996,christopoulos2007}. 
Given the critical role lasers play in the realization of integrated photonics, such phenomena are the focus of considerable attention~\cite{fraser2016}.

More recently, modern devices are using significant dispersion engineering to improve their characteristics or gain access to fundamental physics.
Periodic micropillar structures have been used to demonstrate condensation in gap solitons~\cite{tanese.natcomm.2013}, localisation in flat bands in various lattice geometries~\cite{baboux.PhysRevLett.116.066402,jacqmin.PhysRevLett.112.116402,harder.acsphotonics.2021}, and KPZ physics~\cite{fontaine.nature.2021}.
Topologically robust exciton-polariton lasers~\cite{harari2018} have been demonstrated in one~\cite{solnyshkov2016,st-jean2017} and two~\cite{klembt2018,wu2023} dimensions using patterned structures. 
Problems in patterning polariton microcavities are mostly related to their deep etching through the whole structure and the need to avoid exciton quenching. 
This has usually led to super-wavelength periodicity, small Brillouin zones and small gap openings.
More recently, etched waveguides have allowed to explore the physics of polariton condensation in confined BiC states using subwavelength gratings~\cite{ardizzone2022,hsu2016,gianfrateReconfigurableQuantumFluid2024,riminucciPolaritonCondensationGapConfined2023,trypogeorgos2024emergingsupersoliditypolaritoncondensate}
As such, there remains a compelling need to delve deeper into the condensation dynamics within specific nanostructures where the energy-momentum landscape is more complex with a much better linewidth-to-gap ratio~\cite{fontaine2022,jacqmin2014a,klembt2018,suchomel2018,solnyshkov2022}.

Here, we have deliberately designed such a structure where a single adiabatic mode shows multimode condensation, varying topology and effective mass, creating small energy gaps.
The rich energy-momentum landscape of this two-dimensional patterned waveguide leads to the presence of both symmetry-protected BiCs and accidental polarization vortices coexisting on the polaritonic dispersion.
The resulting threshold behaviour and population dynamics of the condensed states show a competition for the gain reservoir and elucidates the non-trivial role that topology, confinement, and strong-coupling play.

\section{System and band structure} 
\begin{figure*}[htb]
\centering
\includegraphics[width=0.98\textwidth]{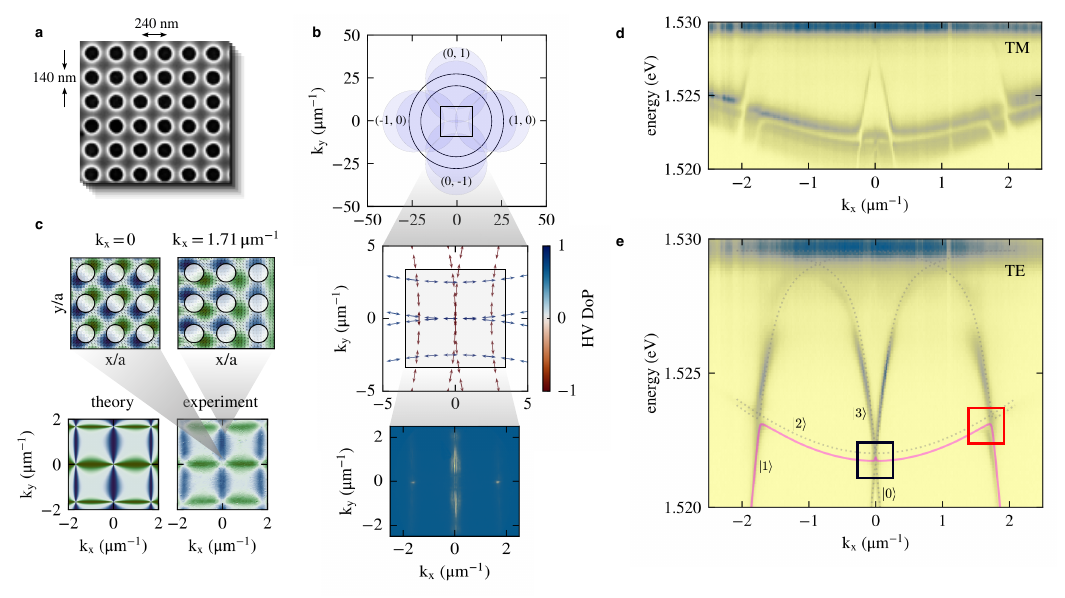}
\caption{System description. a. Scanning electron microscope image showing the patterned surface of the waveguide b. Top panel, schematic representing the reciprocal space geometry at a given energy composed by 8 cones emanating from different points in reciprocal space. The TEM$_0$ (TEM$_1$) modes emanate from the inner (outer) ring. Middle panel, zoom in of the $k = 0$ region; the coloured arrows portray the polarization pattern resulting from the TE nature of the modes. Lower panel, reciprocal space vertically-polarised PL above condensation threshold. c. Bottom, theoretical and experimental reciprocal space S1 degree of polarization integrated in energy. Top, calculated real space field pattern for the $\ket{1,k=0}$ and $\ket{1,k=k_r}$ state respectively d. Energy resolved vertically and e. horizontally-polarised, non resonantly excited, reciprocal space PL below condensation threshold for $k_y=0$. The dotted lines show the theoretical dispersion and the solid magenta line indicates the `moustache' mode. The red and blue boxed regions in e. are where the two condensates appear, as shown in \cref{fig:2}a.}
\label{fig:1}
\end{figure*}

We use an active GaAs/AlGaAs two-dimensional waveguide in the strong-coupling regime with a sub-wavelength, square lattice.
The lattice constant is $a=\qty{240}{nm}$ and the diameter of the circular holes is \qty{140}{nm}, as shown in \cref{fig:1}a (see \cref{sec:fabrication} for more details).
The lattice introduces a diffractive mechanism responsible for coupling counter-propagating guided modes; at the same time, it folds them into a Brillouin zone within the light cone and, as such, makes them amenable to imaging with spectroscopic techniques.

The dispersion of our system, for a given polarisation, can be qualitatively understood by a simple construction involving four sets of Dirac cones displaced in the conjugate plane by a reciprocal lattice vector $1/a$ (\cref{fig:1}b, top).
Each set of cones corresponds to TEM$_0$ or TEM$_1$ pairs of modes whose energy offset is set by the transverse area of the waveguide.
The TEM$_0$ (TEM$_1$) modes originate from a ring in reciprocal space of radius $1/a$ ($1/a + k_r$).
This effective model gives an accurate representation of the dispersion and is in agreement with numerical calculations using both finite elements and rigorous coupled-wave expansion (see \cref{sec:numerics}).

Considering the reciprocal space geometry for the TE modes, whose polarisation is tangent to the circles for any iso-energetic plane, we identify nine points amongst the conical dispersions, corresponding to the distinct, unordered 2-subsets of $\{0, \pm k_r\}$ (\cref{fig:1}b, middle) where orthogonal polarisations cross.
The lower-energy modes of these crossings are the ones predominantly populated via off-resonant pumping due to the confinement and their negative mass, as shown in a vertically-polarised image of the reciprocal space (\cref{fig:1}b, bottom).

When the system maintains a discrete reflection or rotational symmetry, as is the case of a one- or two-dimensional lattice, the odd modes that become symmetry-protected BiCs at the $\Gamma$-point are rigorously given by group theory~\cite{salerno2022loss}.
In this case a polarisation vortex forms at the Dirac cone, which can, however, also form at the off-$\Gamma$ crossing points, as demonstrated by the measured and calculated $S_1$ polarisation vector (\cref{fig:1}c bottom).
The field distribution in all these points is odd with respect to the lattice sites due to the cross-polarised coupled modes.

Since our lattice has C$_4$ symmetry, in the following we will focus our description on the $\Gamma\to X$ direction along $k_x$.
Along this direction, the system comprises two pairs of counter-propagating linear photonic modes TEM$_0^{(\pm 1,0)}$ and TEM$_1^{(\pm 1,0)}$, and two degenerate parabolic modes TEM$_0^{(0,\pm 1)}$ (where the upper indices denote the direction in the reciprocal space, see \cref{fig:1}b). 
The diffractive coupling stemming from the patterning is responsible for the opening of multiple gaps in the dispersion at the $\Gamma$-point but also at finite in-plane wavevector and in lifting the degeneracy of the parabolic modes.
This leads the appearance of two BiC states at the middle and top branches of the $\Gamma$-point where condensation is expected due to the low losses and negative mass of these states~\cite{riminucciPolaritonCondensationGapConfined2023,gianfrateReconfigurableQuantumFluid2024}. 
Moreover two additional gaps open at $k_x=\pm k_r$, where TEM$_0^{(0,\pm 1)}$ and TEM$_1^{(\pm 1, 0)}$ modes cross.

The lattice constant $a$ has been chosen so that the Dirac cone appearing at the crossing of the counter-propagating TEM$_1^{(\pm 1, 0)}$ modes is \qty{12}{meV} red-detuned from the heavy-hole excitonic transition at $E_X=\qty{1531}{meV}$.
The photonic modes are coupled through the grating with Rabi frequencies $\approx\qty{0.15}{meV}$ and all modes are strongly coupled to the exciton with a Rabi frequency of \qty{5.2}{meV} giving rise to the polaritonic eigenstates of our system which we label as $\ket{n}$, where $n\in \mathbb Z$, in order of increasing energy.

Imaging the reciprocal-space photoluminescence (PL) in vertical or horizontal polarisation allows us to gain more insight into the dispersion.
Collecting the vertically-polarised component (\cref{fig:1}d) the PL is dominated by eigenstates that belong to the TM bands. 
The counter-propagating TEM$_1^{(\pm 1, 0)}$ modes cross the parabolic mode at $k_r \equiv k_x \approx \qty{1.74}{\mu m ^{-1}}$, and cross each other roughly at the exciton energy.
For symmetry reasons~\cite{rosenbergPhysRevB.93.195151,burstein2012confined} the TM-like modes are almost completely decoupled from the exciton allowing us to hereafter limit the discussion to TE-like modes only.
The polarisation of the modes rotates with energy which leads to the modes appearing in parts as missing PL; see \cref{sec:numerics} for details.

On the contrary, in horizontal polarization, the TEM$_1^{(0, \pm 1)}$ modes emerge (\cref{fig:1}e), showing an anticrossing with the HH exciton. 
At the $\Gamma$-point a four-level structure emerges with gap of \qty{0.4}{meV} between $\ket{0}$ and the degenerate $\ket{2}$, $\ket{3}$.
Two distinct BiCs with a infinitesimal linewidth emerge at $\ket{0}$ and $\ket{1}$.
At $k_r$, $\ket{1}$ and $\ket{2}$ anticross forming a gap of \qty{1}{meV}.
For near-zero wavevectors, $\ket{1}$ has a negative effective mass of \qty{0.28}{eV s^2/\micro m^2}, which changes to a positive one for $k > \qty{0.2}{\micro m^{-1}}$ and becomes negative again with a value of \qty{1.22}{eV s^2/\micro m^2} at $k_r$, which gives the mode its characteristic `moustache' shape.

\subsection{Effective Hamiltonian}

Working with an effective Hamiltonian based on coupled-modes leads to a comprehensive understanding of the main factors contributing to the observed dispersion shape. 
For simplicity, we will focus only on the TE modes, with purely real frequencies, but losses can be included~\cite{zanotti2022}. 
The light-cone dispersion is
\begin{equation}
    \mathcal{E}(k_x, k_y)_{m,n} = \hbar c_S\sqrt{(k_x \pm m G_x)^2 + (k_y \pm nG_y)^2} + \epsilon,
\end{equation}
where the in-plane wave vectors are shifted by integer multiples of the reciprocal lattice vectors $\mathbf{G_x}= \frac{2 \pi}{a}\mathbf{\hat{x}}$ and $\mathbf{G}_y= \frac{2 \pi}{a}\mathbf{\hat{y}}$, where $a = \qty{241}{nm}$. 
The speed of light in the waveguide $c_S$ provides the light-cone slope, and the offset $\epsilon= \qty{1.522}{eV}$ fixes the energy of the $\Gamma$-point. 
We consider two modes from the waveguide, $\mathcal{E}^{(s)}_{m,n}$ and $\mathcal{E}^{(p)}_{m,n}$ separated by an energy offset determined by the thickness of the waveguide, which for our numerical calculations is $\Delta= \qty{641}{meV}$.
These waveguide modes are coupled to the patterned square array via four sets of light cones located at $(m,n)=\lbrace(+1,0),(-1,0),(0,+1),(0,-1)\rbrace$. 
The behavior of our system can be effectively depicted by the dispersion pattern resulting from the presence of eight of such translated cones, as illustrated in \cref{fig:1} of the main text. 
The linear photonic modes are the $\mathcal{E}_{\pm 1,0}$, while the parabolic modes are the $\mathcal{E}_{0,\pm 1}$. 
The effective $9 \times 9$ Hamiltonian, including the coupling with the exciton, is the following:
\begin{widetext}
\begin{equation}
H= \left( {\begin{array}{*{20}{c}}
{{\mathcal{E}_{1,0}^{(s)}}}&{{\Omega_s}}&{-\Omega_s/2}&{-\Omega_s/2}&{{\Omega_{sp}}}&{\Omega_{sp}}&{\Omega_{sp}}&{\Omega_{sp}}&\Omega \\
{{\Omega_s}}&{{\mathcal{E}_{-1,0}^{(s)}}}&{-\Omega_s/2}&{-\Omega_s/2}&{\Omega_{sp}}&{{\Omega_{sp}}}&{\Omega_{sp}}&{\Omega_{sp}}&\Omega \\
{-\Omega_s/2}&{-\Omega_s/2}&{{\mathcal{E}_{0,1}^{(s)}}}&{{\Omega_s}}&{\Omega_{sp}}&{\Omega_{sp}}&{{\Omega_{sp}}}&{\Omega_{sp}}&\Omega \\
{-\Omega_s/2}&{-\Omega_s/2}&{{\Omega_s}}&{{\mathcal{E}_{0,-1}^{(s)}}}&{\Omega_{sp}}&{\Omega_{sp}}&{\Omega_{sp}}&{{\Omega_{sp}}}&\Omega \\
{{\Omega_{sp}}}&{\Omega_{sp}}&{\Omega_{sp}}&{\Omega_{sp}}&{{\mathcal{E}_{1,0}^{(p)}}}&{{\Omega_p}}&{\Omega_p}&{\Omega_p}&\Omega \\
{\Omega_{sp}}&{{\Omega_{sp}}}&{\Omega_{sp}}&{\Omega_{sp}}&{{\Omega_p}}&{{\mathcal{E}_{-1,0}^{(p)}}}&{\Omega_p}&{\Omega_p}&\Omega \\
{\Omega_{sp}}&{\Omega_{sp}}&{\Omega_{sp}}&{\Omega_{sp}}&{\Omega_{p}}&{\Omega_{p}}&{{\mathcal{E}_{0,1}^{(p)}}}&{{\Omega_p}}&\Omega \\
{\Omega_{sp}}&{\Omega_{sp}}&{\Omega_{sp}}&{{\Omega_{sp}}}&{\Omega_{p}}&{\Omega_{p}}&{{\Omega_{p}}}&{{\mathcal{E}_{0,-1}^{(p)}}}&\Omega \\
\Omega &\Omega &\Omega &\Omega &\Omega &\Omega &\Omega &\Omega &{{E_X}}
\end{array}} \right)
\end{equation}
\end{widetext}
where $E_X$ represents the energy of the exciton, $\Omega$ represents the Rabi coupling, and $\Omega_{s}$, $\Omega_{p}$, $\Omega_{sp}$ represent the different couplings between the TE modes. 
The eigenvalues of this Hamiltonian are subsequently overlayed in \cref{fig:1}de, as two-dimensional cross-sections at $k_y=0$. T
he parameters used to fit the experimental dispersions are $\Omega_s = \qty{-0.2}{meV}$, $\Omega_p = \qty{0.1}{meV}$, $\Omega_{sp} = \qty{0.2}{meV}$, $\Omega = \qty{5.2}{meV}$, $E_X = \qty{1534}{meV}$.

The electric field $E$ in the $x$-$y$ plane, shown in in \cref{fig:1}c, can be reconstructed from the basis of plane waves originating from the $\Gamma$ points at $\pm \mathbf{G}_x$ and $\pm \mathbf{G}_y$~\cite{liang2011three} as:

\begin{equation}
    \begin{split}
        E_x(x,y) &= \psi_y^{(+)} e^{i[k_x x +(k_y+G_y)y]} + \psi_y^{(-)} e^{i[k_x x + (k_y-G_y)y]}\\
        E_y(x,y) &= \psi_x^{(+)} e^{i[(k_x+G_x)x-k_y y]} + \psi_x^{(-)} e^{i[(k_x-G_x)x + k_y y]}
        \label{eq:Efields}
    \end{split}
\end{equation}
where $\psi_{x,y}^{(\pm)}$ are the Bloch-waves coefficients of the mode satisfying $H\psi = E \psi$, and the $\pm$ indicates the propagation direction.

\section{Quasi-steady state} 
\begin{figure}[htbp]
\centering
\includegraphics[width=0.99\linewidth]{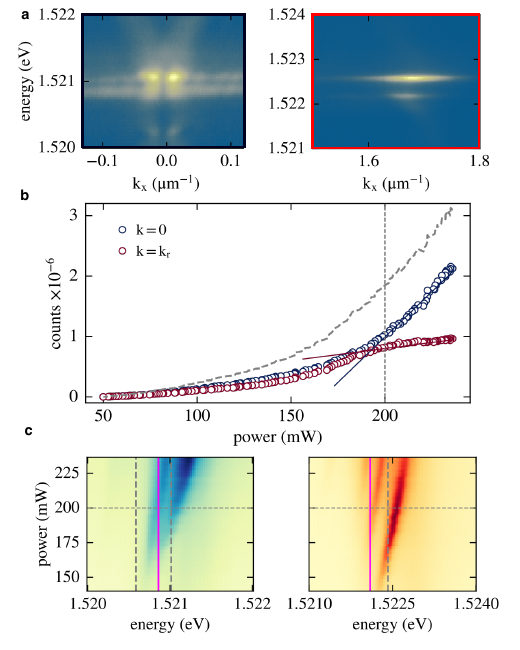}
\caption{a. Horizontally-polarised reciprocal space PL above condensation threshold showing a BEC at $k_x = 0$ and at $k_x = k_r$ corresponding to the marked regions in \cref{fig:1}d.
b. Integrated polariton emission around $k_x = 0$ (blue) and at $k_x = k_r$ (red) as a function of the applied power, showing the threshold behaviour. 
The gray dashed line is the sum of the two curves.
c. Reciprocal space PL distribution for $\ket{1, 0}$ (blue) and $\ket{1,k_r}$ (red) showing the blue-shift of the states at $\Gamma$ and off-$\Gamma$, where two and three states are seen respectively.
The vertical lines are the energies of the modes at $k=0$ from \cref{fig:1}c.
}
\label{fig:2}
\end{figure}

The system was excited via continuous wave, non-resonant pumping, employing a \qty{22}{\micro m}-diameter laser spot, blue-detuned with respect to the excitonic transition. 
The photogenerated electron-hole pairs rapidly relax creating an excitonic reservoir spatially localised below the pump spot. 
Increasing the population, two distinct BECs are formed on the moustache mode at $k=0$ and $k=\pm k_r$ with multiple excited states in the respective gaps (see \cref{fig:2}a.
Due to the small gaps present in the system the two BECs have similar Hopfield coefficients; 0.29 for $k=0$ and 0.35 for $k_r$.
This gives us a unique opportunity to study condensation on a single mode, where the two coupled BECs differ mainly in their topology and effective mass.

The injected exciton-polaritons relax and condense at $k_r$ and at $k=0$ as the pump rate increases notwithstanding the fact that these two points have different topology.
This unexpected behaviour is related to the fact that although a BiC has vanishing linewidth, condensation in gap-confined modes is largely driven by the the negative effective mass~\cite{riminucciPolaritonCondensationGapConfined2023} and coupling to the exciton reservoir, see \cref{sec:vortex}).
At the same time, the momentum spread of the BEC at $k=0$ markedly decreases, indicating that a phase order is formed over a macroscopic distance of \qty{40}{\micro m}. 

The threshold curves for the condensation of each mode are calculated by integrating the PL in energy-resolved measurements. 
As shown in \cref{fig:2}b, the two BECs display similar thresholds after which the populations grow inversely proportional to the mode losses ($\gamma_0 < \gamma_r$). 
The condensation threshold is at roughly $P=\qty{160}{mW}$. 
Figure~\ref{fig:2}a shows the coherent emission that originates from the $\Gamma$ and off-$\Gamma$ points, indicating that the condensation threshold is reached. 
Due to the confinement imposed by the laser, it is the first (second) excited state that condenses at the $\Gamma$ (off-$\Gamma$) point, while the lower-energy states are blue-shifted outside of the gap.
Increasing the pumping power, the populations of the two BECs diverge around $P=\qty{180}{mW}$. 
Thereafter, the $k=0$ BEC becomes dominant.
As pumping power increases, both BECs are blue-shifted and their linewidth initially decreases and then remains nearly constant (\cref{fig:2}c). 

\subsection{Coherence of multi-state systems}

Measuring the spatial coherence of a polaritonic system is typically done using a zero-delay, balanced Mach-Zender interferometer where the image is reflected about the centre in one arm.
For single-mode systems, this produces a stationary fringe pattern whose visibility is proportional to $g^{(1)}(x - x^\prime)$.
In our two-dimensional system, condensation takes place at different energies which introduces time-dependent terms in the coherence function. 

\begin{figure*}[tbp]
\centering
\includegraphics[width=\linewidth]{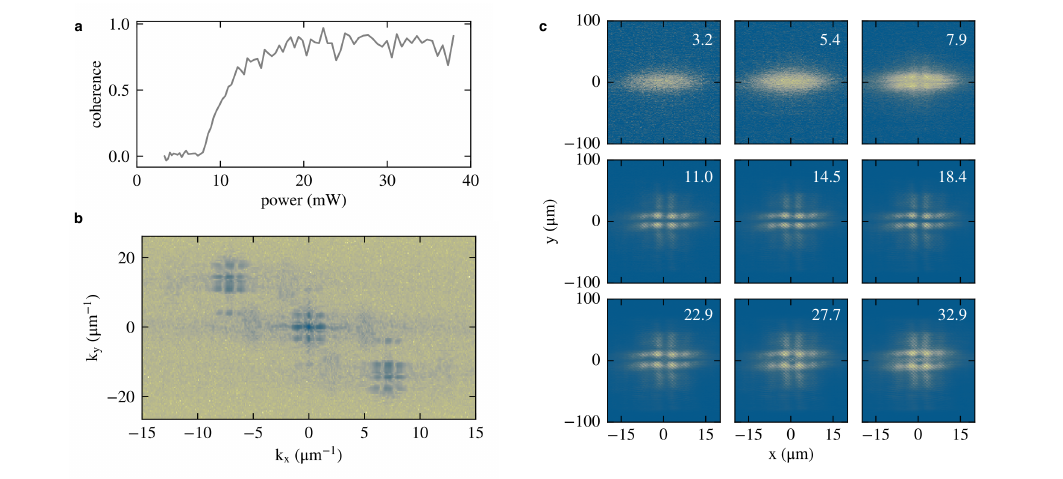}
\caption{Coherence. a. Growth of integrated coherence with applied pump power. The coherence saturates at around $p=\qty{20}{mW}$ when the size of the BEC becomes comparable to the size of the pump laser. b. Reciprocal space of a typical interferometric image. c. Coordinate space interferograms for different powers, showing how different polaritonic modes coexist along the $x$ and $y$ axes.}
\label{fig:s5}
\end{figure*}

The input field consists of two components $E(x; E_1, E_2) = E_1(x) e^{-i\omega_1t} + 2E_2(x) e^{-i\omega_2t} \cos{k_rx}$, where $\hbar\omega_{1,2}$ are the energies of the two condensates and $\omega_2 > \omega_1$.
To calculate $g^{(1)}(x - x^\prime; t)$ we consider the superposition of $E(x; E_1, E_2)$ and $E(x; E_3, E_4) e^{-ik_1x}$, where $k_1$ is the wavevector of the fringe pattern, corresponding to the relative angle of incidence of the two images from the interferometer at the imaging plane.
It is evident that $I(x) = |E(x; E_1, E_2) + E(x; E_3, E_4) e^{-ik_1x}|^2$ will contain terms that oscillate at frequencies $\omega_1$, $\omega_2$, $\omega_1 + \omega_2$, $\omega_2 - \omega_1$.
Since the energy difference between the two condensates is about \qty{1}{meV}, the longest period of oscillation $T = 2\pi(\omega_2 - \omega_1)^{-1}$ corresponds to \qty{4}{ps} which is much smaller than the image acquisition time.
As such we consider the time-integrated intensity of the above superposition
\begin{equation}
\begin{split}
    I(x) & = \frac1T\int_0^T{I(x;t) dt} = \\
    & = 2E_{1}^{2} + 8 E_{2}^{2} \cos^{2}{\left(k_{r} x \right)}
    + 2(E_{1}^2 + 2E_2^2) \cos{\left(k_{1} x \right)} \\
    & + 2 E_{2}^2 \left(\cos{\left(k_{1} x - 2 k_{r} x \right)} + \cos{\left(k_{1} x + 2 k_{r} x \right)}\right),
\end{split}
\end{equation}
where the spatial dependence has been dropped from the fields for clarity $E_i(x)\equiv E_i$ and we have assumed the interferometer is balanced $E_1=E3$ and $E_2=E_4$.
It is possible to extract the coherence of each condensate by considering the individual reciprocal components of the above, $k_r$, $k_1 - k_r$, $k_1$, $k_1 + k_r$ and defining a visibility associated with the respective dc terms.
In our experiment however, the components arising from the interference of the two wavevectors $k_1\pm k_2$ are negligible for the smaller powers.
As such we chose to consider a dc cutoff slightly higher than $k_r$ and a bandwidth larger than $2k_r$ for the fringe component, effectively calculating the total coherence of both condensates.

The transition from spontaneous emission to the condensate was confirmed by a coherence measure, as shown in \cref{fig:s5}a.
We employed a Michelson interferometer for measuring the spatial coherence of the BEC, using a retroreflector to symmetrically flip one of the two images before interfering onto the imaging plane. 

From Fourier analysis of the interferogram (\cref{fig:s5}b) we extracted the energy-integrated, first-order total coherence function $g^{(1)}(x-x^\prime)$ of both BECs that saturates to a finite-size limited value at around $P=\qty{20}{mW}$.
Typical examples of the interferograms can be seen in of \cref{fig:s5}c.

\section{Dynamics}
By switching to pulsed excitation, we employed energy-time resolved measurements of the reciprocal space PL, using a streak camera with a resolution of \qty{1}{ps}, in order to investigate the temporal population dynamics of the multimode polariton condensate.
To get the full dynamics of the dispersion, we imaged the BEC density for each value of $k_x$ from \qtyrange{-2}{0.3}{\micro m^{-1}}.
Referencing $t=0$ to the arrival of the excitation pulse, we can reconstruct the polaritonic dispersion at any later time.

\begin{figure}[htbp]
 \centering
 \includegraphics[width=\columnwidth]{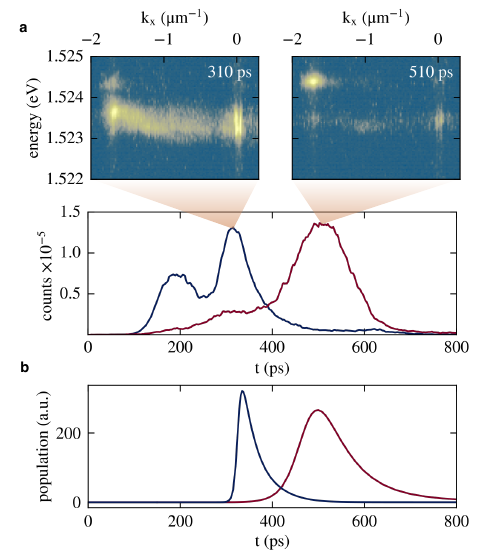}
 \caption{
 Dynamic evolution of the system. a. Snapshots of the dispersion population at the times of peak emission for the two condensates (top). The dynamical behaviour of the $k=0$ (blue) and $k=k_r$ (red) condensate (bottom). The resonances before the emission peaks correspond to states that traverse the gap without condensing till eventually one the first excited state condenses at the $\Gamma$-point at \qty{310}{ps} and the second excited state at the off-$\Gamma$ point at \qty{510}{ps}. b. Population dynamics from the rate equations \cref{eq:rate} show only the condensed states.
 }
 \label{fig:3}
\end{figure}

Figure~\ref{fig:3} shows the time evolution of the condensation of the two BECs for an excitation power corresponding to a power of about \qty{200}{mW}, marked with a vertical grey dashed line in \cref{fig:2}b.
The two BECs are characterised by a lifetime of around \qty{200}{ps} and \qty{90}{ps}, at $k=0,$ and $k=k_r$ respectively; the lifetime was extracted from an exponential fit of the decaying tails of population in \cref{fig:3}.

Surprisingly, we observed rich and complex dynamics, previously only seen in spatially separated BECs pinned on structured potentials arising from defects or patterning~\cite{PhysRevLett.112.113602}.
In our system, a similar situation arises naturally due to the BECs being strongly localised and confined under the same laser spot.
Two main features of the dynamical behaviour are most prominent: a. the peak emission from the condensate at $k=k_r$ is delayed by approximately \qty{200}{ps} from the peak emission from $k=0$ and b. other states traverse both energy gaps at earlier times without going past the condensation threshold.

Although the two BECs show roughly the same power threshold, the $k=0$ mode condenses \qty{200}{ps} earlier.
These time-delayed dynamics are partly due to gain competition for the same localised excitonic reservoir at the growth stage of the BEC population and the relaxation of particles from large to small wavevectors; this can be straightforwardly modelled via coupled Boltzmann rate equations including both an active and inactive exciton reservoir~\cite{PhysRevB.98.195312}.
We obtain a similar coupling rate to the reservoir for both BECs and a smaller coupling rate between them where polaritons flow from $k_r$ to $k=0$.
The non-trivial topology of the BiC, which manifests as an increased lifetime, leads to it condensing first in time, although curiously we do not observe similar behaviour in power for which they both show a similar threshold.

The individual condensate dynamics are intricate in themselves, aside for the initial delay dynamics.
Condensation does not take place for the ground state but for the first (second) excited state for $k=0$ ($k=k_r$) due to the size of the energy gaps that is comparable with the blueshift; the gap at $k=0$ is \qty{270}{\mu eV} and at $k=k_r$ \qty{320}{\mu eV}.
The first states reach their peak amplitude at around \qty{200}{ps} both at $k_r$ and $k=0$, however the emission from the $\Gamma$-point is almost an order of magnitude brighter.
Similarly, the first excited states reach their peak emission around \qty{310}{ps}, at which point the state at the $\Gamma$-point condenses and as such saturates the gain from the reservoir.
This saturation effective delays the condensation of the off-$\Gamma$ mode by \qty{200}{ps}, at which point the second excited state that traverses the gap finally condenses.
The relative peak amplitudes of the emission at \qtylist{202;310;510}{ps} are in agreement with the relative amplitudes of the different states along the \qty{200}{mW} horizontal lines shown spectroscopically in \cref{fig:2}c for continuous-wave excitation.

\subsection{Rate equations}

We model the dynamical formation of the two BECs with a simple two-mode rate equation model~\cite{PhysRevB.98.195312,PhysRevLett.112.113602,opala2018}. 
Such models are effectively zero-dimensional Gross-Pitaevskii equations and as such cannot capture trapped mode dynamics, however they accurately model the competition for the reservoir.

\begin{subequations}
\label{eq:rate}
\begin{align}
    \frac{dn_{c1}}{dt} &= R_1 n_r n_{c1} - \gamma_{c1} n_{c1} + R_{12}n_{c2}n_{c1}, \\
    \frac{dn_{c2}}{dt} &= R_2 n_r n_{c2} - \gamma_{c2} n_{c2}, \\
    \frac{dn_{r}}{dt} &= \kappa n_i^2 - \gamma_r n_r - R_1 n_r n_{c1} - R_2 n_r n_{c2},\\
    \frac{dn_{i}}{dt} &= P(t) - \kappa n_i^2 - \gamma_i n_i.
\end{align}
\end{subequations}

The first two equations describe the dynamics of the polariton BEC densities $n_{c1,2}$ that are coupled to the active excitonic reservoir via coupling constants $R_{1,2}$ and to the environment via $\gamma_{c1,2}$.
Polaritons from the mode at $k=k_r$ can further relax onto the $k=0$ mode at a rate $R_{12}>0$.
The nonlinearity $\kappa$ comes from the inactive excitonic reservoir.
The active and inactive reservoirs dissipate at a rate of $\gamma_r$ and $\gamma_i$ respectively.
Excitonic population is generated via a gaussian-shaped pulse $P(t)$.

\Cref{fig:3}b shows the dynamics of both reservoirs and modes commencing with a \qty{100}{fs} pulse.
The initial populations are $n_{c1,2}(t=0)=\num{1e-5}$, $n_r=0$, and $n_i=\num{1e3}$.
The values of the other parameters are $R_1=\num{4.1e-4}$, $R_2=\num{5.4e-4}$, $R_{12}=\num{9.8e-5}$, $1/\gamma_{c1}=118.3$, $1/\gamma_{c2}=61.7$, $\kappa=\num{8.9e-6}$, $1/\gamma_r=810$, and $1/\gamma_i=2000$.
The arrival of the pulse generates a number of exciton in the inactive reservoir that slowly decays.
The active reservoir starts filling up at a rate $\kappa$, up till the moment when bosonisation takes place at the $k=0$ condensate.
The active reservoir starts refilling as soon as the condensate starts decaying, which allows for the condensate at $k=k_r$ to bosonise \qty{200}{ps} later.

\section{Discussion and conclusions} 
We have shown how engineering the band structure of a polaritonic waveguide via periodic patterning leads to the formation of multiple condensates at different energies and momenta.
This gave us a unique testbed to study the non-trivial behaviour of the polaritons in our system that arises from an interplay of topology, negative effective mass, and confinement. 
Our system is capable of sustaining long-lived polariton condensates, which is a crucial characteristic for potential practical applications of integrated polariton-based devices.
Both condensates appeared on a single polaritonic mode at near-zero group velocity and negative effective mass.
Surprisingly, although their topology is radically different, since the $\Gamma$-point condensate is on a BiC state, the threshold power required was comparable and the difference in linewidth is mainly evident by the different $1/\gamma$ growth rate of the condensed population.

In addition, the condensation dynamics of exciton-polaritons in such two-dimensional square lattice waveguides shows rich phenomenology providing valuable insights into the underlying physics.
Notably, we observed the appearance of two symmetric exciton-polariton condensates at similar power threshold but distinct times; given that the coupling to the excitonic reservoir is comparable for both condensates, this behaviour is connected to their different topology and has implications about the spatial cross-coherence of the system.
The self-confining effect of the negative mass BECs and small energy gaps allowed us to observe interesting dynamics, previously only measured in BECs pinned on structured potentials.
Multiple states traverse the gap without going past the condensation threshold.
Since states at different wavevectors compete for the same reservoir, as long as a state condenses it saturates the reservoir and delays condensation at the other wavevector by a time comparable to its lifetime.
Understanding how to control these dynamics is a central tenet of modern polaritonic devices; our work establishes the fundamental behaviour and paves the way for future studies of photonic crystal polariton waveguides and similar systems with engineered bandstructures.

\vspace{0.3cm}
\acknowledgments{We thank S. Zanotti and D. Gerace for fruitful discussions, and P. Cazzato for technical support. 
This project was funded by PNRR MUR project: `National Quantum Science and Technology Institute' - NQSTI (PE0000023);
PNRR MUR project: ‘Integrated Infrastructure Initiative in Photonic and Quantum Sciences’ - I-PHOQS (IR0000016);
Quantum Optical Networks based on Exciton-polaritons - (Q-ONE) funding from the HORIZON-EIC-2022-PATHFINDER CHALLENGES EU programme under grant agreement No. 101115575;
Neuromorphic Polariton Accelerator - (PolArt) funding from the Horizon-EIC-2023-Pathfinder Open EU programme under grant agreement No.  101130304;
the project ``Hardware implementation of a polariton neural network for neuromorphic computing'' – Joint Bilateral Agreement CNR-RFBR (Russian Foundation for Basic Research) – Triennal Program 2021–2023;
the MAECI project ``Novel photonic platform for neuromorphic computing'', Joint Bilateral Project Italia - Polonia 2022-2023;  
the PRIN project ``QNoRM: A quantum neuromorphic recognition machine of quantum states'' - (grant 20229J8Z4P).
Views and opinions expressed are however those of the author(s) only and do not necessarily reflect those of the European Union or European Innovation Council and SMEs Executive Agency (EISMEA). Neither the European Union nor the granting authority can be held responsible for them.
G.S. was funded by the Research Council of Finland through project n. 13354165.
}

\appendix

\section{Experimental techniques}

\subsection{Sample fabrication}
\label{sec:fabrication}

Our sample is a planar, \qty{500}{nm} AlGaAs slab, embedding 12 GaAs, \qty{20}{nm}-thick, quantum wells and closed by a \qty{10}{nm} capping GaAs layer. 
The structure is grown on top of a couple of heavily aluminium doped layers with a total thickness of \qty{550}{nm} which act as a cladding.
By etching the first layers of the waveguide using an array of cylinders with a radius of \qty{70}{nm} and a depth of \qty{140}{nm}, we create a square lattice with a lattice constant $a=\qty{241(1)}{nm}$, covering a \qtyproduct{300 x 300}{\micro m} region.
The sample was fixed with heat-conducting silver paint on a copper holder inside a Montana cryostat at around \qty{4}{K}.

\subsection{Photoluminescence measurements}

The optical setup allows for the acquisition of the integrated emitted spectra, or alternatively the spatial or angle dispersion profile of the emission. 
We performed non-resonant photoluminescence (PL) using the same photographic objective lens, with $NA=0.34$, for both excitation and collection.
The sample was excited out of resonance using a femtosecond pulsed laser (\qty{100}{fs} duration and \qty{80}{MHz} repetition rate) with a center wavelength of \qty{775}{nm}. 
Photons that are released as a result of the radiative recombination of polaritons are emitted at a finite angle $k_{//} = \left( {E/\hbar c} \right)\sin \theta$, where $E$ and $k_{//}$ are the polariton energy and in-plane wavevector. 
The PL is focused on the entry slits of a spectrometer connected to a CCD camera, which is used for real- and reciprocal-space measurements. 
The spectrometer has an energy resolution of about \qty{30}{\micro eV}. 
For all optical measurements, we polarized the emission to either TE or TM polarization.

\subsection{Changing lattice constant}

\begin{figure}[htbp]
\centering
\includegraphics[width=\columnwidth]{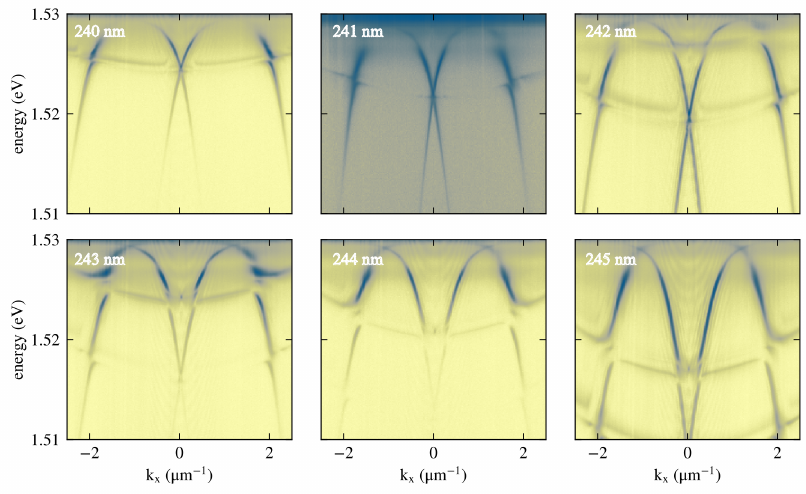}
\caption{Dispersions of different two-dimensional photonic waveguides in order of increasing lattice constant. In the main text we used the \qty{241}{nm} lattice. Both TE and TM parabolic modes are shifting towards lower energies for larger lattice constants.}
\label{fig:s2}
\end{figure}

At \qty{241}{nm} (the lattice constant used in the main text) the parabolic TE$_0$ modes cross at $<\qty{1}{meV}$ below the exciton while TM$_0$ are almost resonant with it.
These parabolic modes cannot be discerned in PL since they are counter-polarised to the rest of the dispersion.
Increasing the lattice constant effectively changes the detuning between the exciton and the coupled propagating photonic modes which shift towards the lower energies.
This eventually leads to the TM parabolic modes being visible below the excitonic line for lattice constants $>\qty{243}{nm}$.
The formation of accidental BICs is evident in these modes which is due to the TE and TM modes being $\pi$ out of phase at a given $k$.
These modes appear bright in PL since they are co-polarised with the rest of the dispersion.
The large gaps that are formed when the TM modes are below the exciton act to disconnect the excitonic reservoir from the photonic modes and inhibit condensation.
A bottleneck region is formed right below the exciton, but the carriers are unable to relax further.

\subsection{Polarised imaging}

In the 1D BiC, where the system is characterized by a $C_2$ symmetry, the condensate polarization is set by the TE propagating modes and results in almost completely linear polarization, except for a small radius of curvature given by the cones that drive the appearance of the characteristic BiC vortex~\cite{ardizzone2022}
In the 2D case, to a first approximation, the condensate appearing at the $\Gamma$-point can be thought of as the superposition of two orthogonal BiCs, both in directionality and polarization, intersecting at the $\Gamma$-point.
\begin{figure}[htbp]
\centering
\includegraphics[width=\columnwidth]{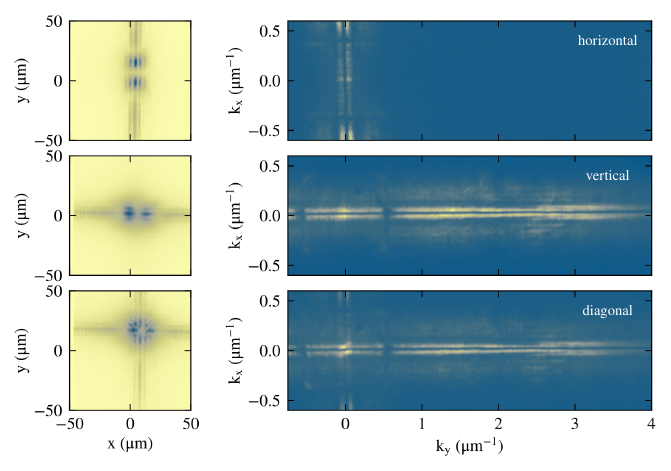}
\caption{Polarised imaging. Coordinate and reciprocal space images for horizontal, vertical, and diagonal polarisation.}
\label{fig:s7}
\end{figure}

Therefore, by selecting one of the two linear polarisations in detection, it is possible to clearly discern between them, untangling the condensate structure in coordinate and conjugate space.
In \cref{fig:s7}, the top (middle) row shows that the BiC envelope is evident along the $y$ ($x$) axis. 
The wavefunction clearly shows three nodes, suggesting that the state reaching the critical density is actually the second excited state of the BiC trap. 
The bottom panel is diagonally polarised and shows effectively a superposition of the two linearly polarised images above.

\section{Energy-resolved degree of polarisation}
\label{sec:vortex}

\begin{figure}[htbp]
\centering
\includegraphics[width=\columnwidth]{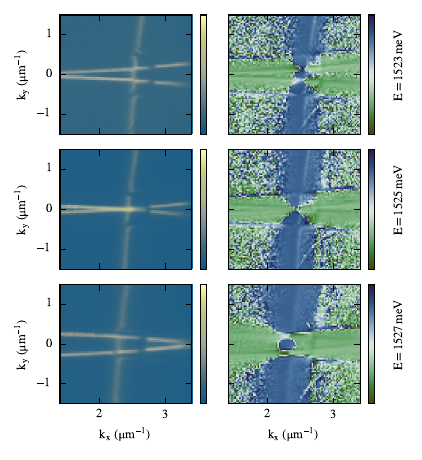}
\caption{Energy-resolved polarisation tomography around $k_r$. Unpolarised luminescence profiles (left column), and equatorial angle of the polarization ellipsoid  $\phi = \arctan(S_1/S_2)$ (right column) for \qtylist{1523;1525;1527}{meV}.}
\label{fig:s8}
\end{figure}

For $k_x = k_r$ the energy-integrated, reciprocal-space photoluminescence shows the clear presence of a polarization vortex (see \cref{fig:1}{c})
However, despite this, the corresponding population profile of condensate shows no signature of the a phase discontinuity, typical of a BiC state, suggesting it is topologically trivial.
The situation becomes clearer in energy-resolved photoluminescence tomography that shows that the polarization vortex at $k_x = k_r$ exists only for a very specific energy of \qty{1525}{meV}.
When the energy is either red or blue detuned the BiC moves towards finite $\pm k_y$. 
This behaviour can be seen in \cref{fig:s8}, both in luminescence and the the polarization ellipsoid angle, $\phi$. 
At an energy of \qty{1523}{meV} the BiC has already moved away from the $k_x$ axis and the condensate lies on a TEM$_1^{(0,\pm 1)}$ mode with trivial topology. 
This happens because, as discussed in \cite{riminucciPolaritonCondensationGapConfined2023}, the condensation in these systems is predominantly facilitated by the local, negative effective mass which leads to a self trapping effect, rather than the  linewidth reduction given by the topological nature of the BiC.

\section{Numerical simulations}
\label{sec:numerics}

\begin{figure*}[htbp]
\centering
\includegraphics[]{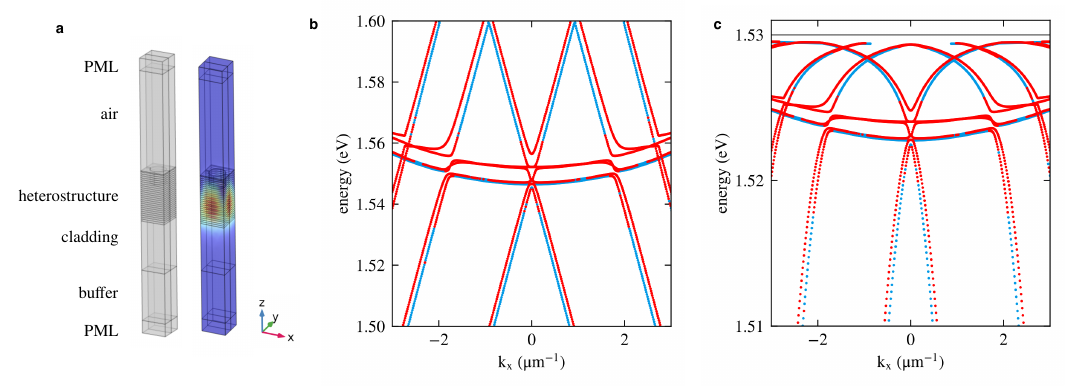}
\caption{COMSOL simulations. a. Unit cell of the square lattice heterostructure; see \cref{sec:fabrication} for the precise composition of the sample. b. Photonic and c. polaritonic dispersions showing TE- and TM-like modes in red and blue. Notice how the polarisation of individual modes is mixed close to the coupling points.}
\label{fig:s3}
\end{figure*}

\subsection{Photonic dispersions}

 To obtain numerically the dispersion, we used a finite element method software (COMSOL), that solves the Helmholtz equations:
\begin{equation}\label{Helm}
    \nabla \times (\nabla \times \mathbf{E}(\mathbf{r}))=k^2 \epsilon_\mathrm{r}(\mathbf{r})\mathbf{E}(\mathbf{r}),
\end{equation}
where $\mathbf{E}(\mathbf{r})$ is the electric field profile, $\epsilon_\mathrm{r}(\mathbf{r})$ the permittivity tensor, and $k$ the wavevector.  
Periodic boundary conditions were used in $x$ and $y$ directions and a symmetric mesh. 
Along the $z$ direction, perfect matched layers (PML) were used to eliminate unphysical reflections due to the finite size of the simulation (see \cref{fig:s3}a). 
We calculated the spatial profiles and the energies of all photonic eigenmodes along the full Brillouin zone. 
Since this is a three-dimensional structure, we differentiate between quasi-TE and quasi-TM photonic modes (shown in \cref{fig:s3}b) by comparing the magnitudes of the electric and magnetic fields along the $z$-axis. 
For quasi-TE modes, the electric field $E_z$ is zero, while for quasi-TM modes, the magnetic field $H_z$ is zero. 
In practice we differentiate the modes by using a physically-informed value as a threshold for the $|E_z|/|H_z|$, since numerical artefacts produce fields that always have a near-zero but finite value.

\subsection{Polaritonic dispersions}

The polaritonic eigenenergies can be obtained by using the photonic eigenergies and coupling them to an effective excitonic resonance~\cite{PhysRev.112.1555,kavokinmicrocavitiesbook}. 
This gives
\begin{equation}\label{strongcoupling}
    E_{pol}= \left( \begin{array}{*{20}{c}}
E_X & \hbar \Omega_R \\ \hbar \Omega_R & E_p(k)
\end{array} \right),
\end{equation}
where $2\hbar \Omega_R=\qty{14}{meV}$ is the Rabi splitting and $E_X=\qty{1530.5}{meV}$ is the energy of the exciton as obtained from the experimental measurements. 
The resulting avoided crossings are shown in \cref{fig:s3}c.

\clearpage

\bibliography{bibliography}

\end{document}